# MEASUREMENTS AND ANALYSIS OF BEAM TRANSFER FUNCTIONS IN THE FERMILAB RECYCLER RING USING THE TRANSVERSE DIGITAL DAMPER SYSTEM*

N. Eddy[#], J. Crisp, M. Hu, Fermi National Accelerator Laboratory, Batavia, IL 60510, U.S.A.

*Abstract*

The primary purpose of the Fermilab Recycler Ring Transverse Digital Damper System is to prevent instabilities due to high phase space densities of the cooled antiproton beam. The system was designed to facilitate Beam Transfer Function measurements using a signal analyzer connected to auxiliary system ports for timing and diagnostic purposes. The Digital Damper System has the capability for both open and closed loop measurements. The Beam Transfer Function provides direct measurement of the machine impedance, and beam and lattice parameters such as betatron tune and chromaticity. An overview of the technique is presented along with analysis and results from open and closed loop measurements in the Fermilab Recycler Ring.

## INTRODUCTION

The Fermilab Recycler Ring is a permanent-magnet based, 8 GeV anti-proton storage and cooling ring. Both stochastic and electron cooling have been employed to increase the phase space density to meet the collider Run-IIb luminosity goal of $2 \times 10^{32} \text{cm}^{-2}\text{s}^{-1}$. To meet and exceed this requirement, the maximally allowable chromaticities, constrained by beam lifetime considerations, do not provide adequate Landau damping at the phase-space densities required by the luminosity goal. The transverse stability threshold has been consistently observed and well understood for the Recycler [1]. A transverse damper system which provides active beam feedback is used to keep the beam stable. The density threshold with operational chromaticities at which instabilities occur has been increased by a factor of greater than 4 with the current broad-band transverse digital damper system.

## TRANSVERSE DAMPER SYSTEM

A basic diagram of the system is shown in Fig. 1. Two pickups in each plane approximately 90° apart in betatron phase are used to simulate the optimum betatron phase advance between pickup and kicker. The feedback system consists of a beam pickup, a filter to reject revolution harmonics, a delay line, power amplifiers, and a kicker. Transverse damping is accomplished by two independent systems acting in the horizontal and vertical planes. The system provides negative feedback from 15KHz to 70Mhz in each plane which covers the first 780 rotation harmonics.

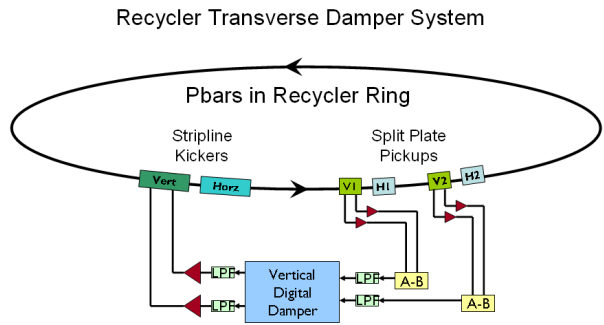

Figure 1: Basic system diagram. Only the vertical plane is shown.

### Digital Filter

The difference signals are input to custom VME boards, one per plane, which digitize the signals at 211MHz and processes them [2]. The board uses a large FPGA to implement digital filtering, gain, and delays as well as diagnostics. Fast digital to analog converters provide the drive signal for the beam kickers. The output rate of 634MHz provides 1.6ns delay resolution on the output signals. The digital filter is shown in Fig. 2. The converters and FPGA are all synchronized to the Recycler 52.8MHz RF. For timing and diagnostic measurements, each board also has one input and ouput channel which can connect to the ports on a signal analyzer.

## BEAM TRANSFER FUNCTION MEASUREMENTS

Beam transfer function (BTF) measurements are performed by driving the beam with an external excitation and measuring the induced signal from a transverse pickup[2]. A signal analyzer is connected to the diagnostic input and output channels as shown in Fig. 2. For calibration the signal is simply looped back so that the effective "extra cable" can be measured.

BTF measurements have been performed with a network analyzer up to 70MHz. These measurements currently lack precision due to the long sweep time required and the fact the Recycler beam is perturbed by stray magnetic fields from the Main Injector ramps which cause orbit and tune shifts. For this reason, a vector signal analyzer (VSA) is used to make precision BTF measurements. The VSA can be gated to take data when the Main Injector is not ramping and also allows fast averaging to improve signal to noise. The VSA is limited to 10MHz so can only be used to measure the first 110 harmonics.

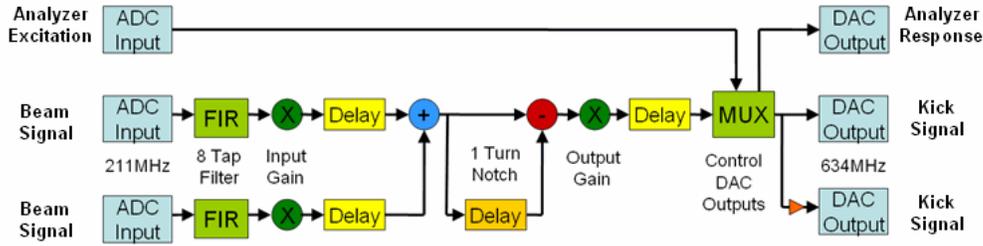

Figure 2: One board per plane processes signals at 211MHz and outputs the feedback kick at 634MHz. A digital MUX is implemented at the output which injecting an excitation signal while opening the loop or keeping the loop closed.

## Open Loop Response

Open loop response is measured by breaking the feedback loop inside the digital logic. In open loop mode, the diagnostic input is driven onto the kicker and the subsequent beam response is sent out the diagnostic output. Open loop measurements are used for initial timing in the system and diagnostics. A network analyzer is used to check the timing out to the 600[th] rotation harmonic. The measured BTFs from the network analyzer are shown in Fig. 3. Open loop measurements can only be made at safe beam densities.

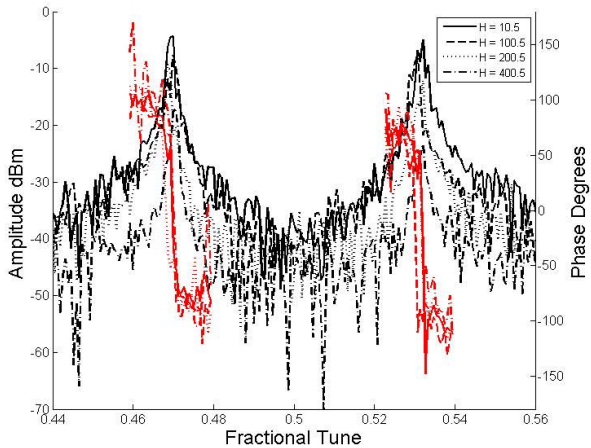

Figure 3: Open loop BTF measurements made at h=10, 100, 200, & 400 using the network analyzer. The signal is rather noisy and shifts due to Main Injector stray fields are evident.

## Closed Loop Response

Closed loop response is measured by adding the excitation signal to the feedback output without breaking the loop. The same sum is also output to the diagnostic port. The closed loop BTF measured with the Vector Signal Analyzer (VSA) are shown in Fig. 4. The closed loop measurements have the advantage that they can be made with the damper system operational. The closed loop response is related to the open loop response by

$$CL_{resp} = \frac{1}{1+OL_{resp}}$$

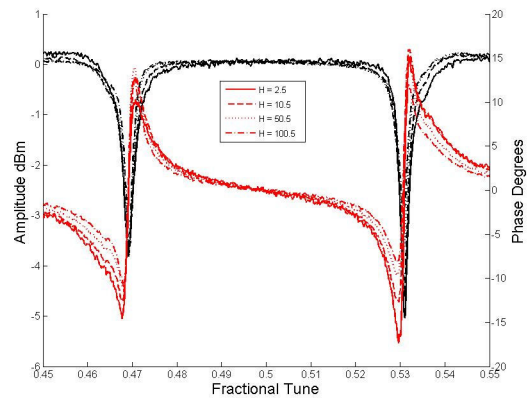

Figure 4: Closed loop response measurements made at h=2,10,50, & 100 using VSA with 64 averages.

The open loop response calculated from the closed loop response in Fig. 4 is shown in Fig. 5.

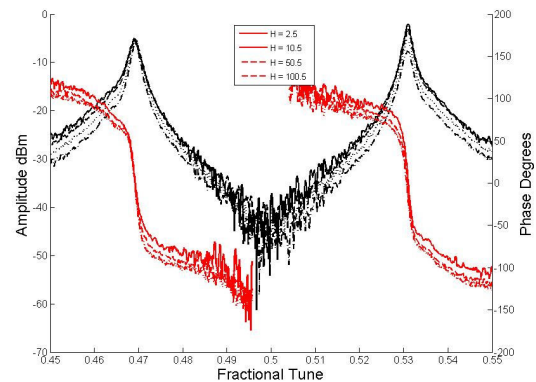

Figure 5: Calculated open loop response measurements made at h=2,10,50, & 100 using VSA with 64 averages.

## Tune Measurement

The easiest beam parameter to measure is the fractional machine tune. Because the Recycler tunes are close to 0.5, one can measure the upper and lower tune harmonics with a 12KHz span centered on N.5 harmonic frequency.

The BTF measurements in Fig. 3, 4, and 5 clearly show the machine tune. In these measurements the upper sideband is on the left and the lower sideband is on the right.

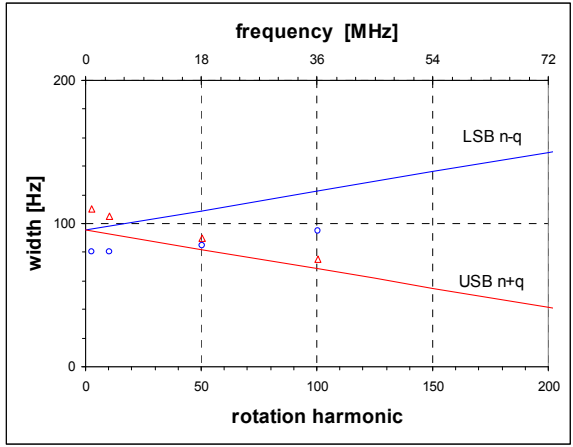

Figure 6: Predicted widths for expected machine parameters (lines) and chromaticity of -3 along with measured widths (points) from the closed loop BTFs.

### Chromaticity Measurement

Chromaticites can be calculated from the measured widths of upper and lower betatron sidebands [4]. Measurements made at different harmonics are shown in Fig. 6 along with the expected chromaticities. The chromaticites measured with this method are consistent with those measured with a different method, in which the RF frequency is slowly varied and the tune shifts are recorded. In the latter method, the Recycler tune shifts vs. changing $\Delta p/p$ showed significant curvature (higher order fields).

### Stability and Impedance

The BTF measurement is related to the transverse machine impedance, $Z_\perp$ [3,4], by

$$Z_\perp = k\left(\frac{I_{b1}}{B_1(\omega)} - \frac{I_{b2}}{B_2(\omega)}\right)\frac{1}{I_{b1} - I_{b2}}$$

Where $B_1(\omega)$ at $B_2(\omega)$ are the BTF measurements at beam currents, $I_{b1}$ and $I_{b2}$, and $k$ is a scale factor which is dependent upon the system gain as well as pickup, kicker, and cable response. The stability diagram is made by plotting $Re(1/B(\omega))$ vs. $Im(1/B(\omega))$ as shown in Fig. 7. The preliminary results show a hint of a shift but it does not exhibit much frequency dependence from the 2$^{nd}$ to 100$^{th}$ rotation harmonic.

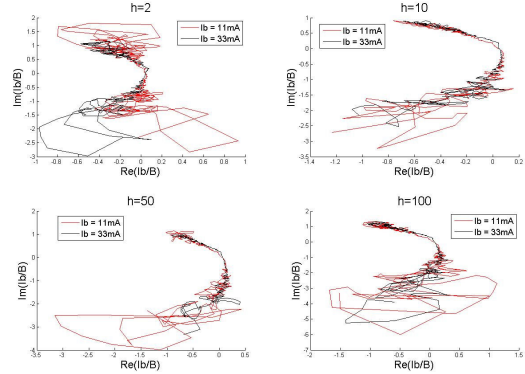

Figure 7: Stability curves for the h=2,10,50, & 100 BTF measurements at beam currents of 11mA and 33mA. A small shift is noted near the origin.

## ACKNOWLEDGEMENTS

The authors would like to thank Bill Ng and Alexey Burov for providing theoretical concepts.